\documentclass[twocolumn,secnumarabic,amssymb, nobibnotes, aps, prd, superscriptaddress]{revtex4-2}

\usepackage{graphicx}
\usepackage{dcolumn}
\usepackage{bm}

\begin{document}

\title{Searches for CE$\nu$NS and Physics beyond the Standard Model\\ using Skipper-CCDs at CONNIE}

\author{Alexis A.\ Aguilar-Arevalo}
\author{Juan Carlos\ D'Olivo}
\author{Y.\ Sarkis}
\affiliation{Instituto de Ciencias Nucleares, Universidad Nacional Autónoma de México,  Circuito Exterior s/n, C.U., CDMX, Mexico}

\author{Nicolas\ Avalos}
\author{Xavier\ Bertou}
\affiliation{Centro Atómico Bariloche and Instituto Balseiro, Comisión Nacional de Energía Atómica (CNEA), Consejo Nacional de Investigaciones Científicas y Técnicas (CONICET), Universidad Nacional de Cuyo (UNCUYO), Av.\ Bustillo~9500, San Carlos de Bariloche, Argentina}

\author{Carla\ Bonifazi}
\email{cbonifazi@unsam.edu.ar}
\affiliation{International Center for Advanced Studies \& Instituto de Ciencias Físicas, ECyT-UNSAM and CONICET, 25 de Mayo y Francia, Buenos Aires, Argentina}
\affiliation{Instituto de Física, Universidade Federal do Rio de Janeiro, Av. Athos da Silveira Ramos, 149, Cidade Universitária, Rio de Janeiro, RJ, Brazil}

\author{Gustavo\ Cancelo}
\author{Juan\ Estrada}
\author{Richard\ Ford}
\author{Kevin\ Kuk}
\author{Andrew\ Lathrop}
\author{Javier\ Tiffenberg}
\author{Sho\ Uemura}
\affiliation{Fermi National Accelerator Laboratory, Batavia, IL, United States}

\author{Brenda A.\ Cervantes-Vergara}
\affiliation{Instituto de Ciencias Nucleares, Universidad Nacional Autónoma de México,  Circuito Exterior s/n, C.U., CDMX, Mexico}
\affiliation{Fermi National Accelerator Laboratory, Batavia, IL, United States}

\author{Claudio\ Chavez}
\affiliation{Fermi National Accelerator Laboratory, Batavia, IL, United States}
\affiliation{Instituto de Inv. en Ing. Eléctrica “Alfredo Desages” (IIIE), Dpto. de Ing. Eléctrica y de Computadoras, CONICET and Universidad Nacional del Sur (UNS), 800 San Andrés Street, Bahía Blanca, Argentina} 
\affiliation{Facultad de Ingeniería, Universidad Nacional de Asunción, Campus de la UNA, San Lorenzo, Paraguay}

\author{Fernando\ Chierchie}
\affiliation{Instituto de Inv. en Ing. Eléctrica “Alfredo Desages” (IIIE), Dpto. de Ing. Eléctrica y de Computadoras, CONICET and Universidad Nacional del Sur (UNS), 800 San Andrés Street, Bahía Blanca, Argentina}

\author{Gustavo\ Coelho Corr\^ea} 
\affiliation{Eletronuclear, Rodovia Procurador Haroldo Fernandes Duarte km 521, Itaorna, Angra dos Reis, RJ, Brazil}

\author{João\ dos Anjos}
\author{Herman P.\ Lima Jr.}
\affiliation{Centro Brasileiro de Pesquisas Físicas, Rua Dr. Xavier Sigaud, 150, Urca,  Rio de Janeiro, RJ, Brazil}

\author{Guillermo\ Fernandez Moroni}
\affiliation{Fermi National Accelerator Laboratory, Batavia, IL, United States}
\affiliation{Instituto de Inv. en Ing. Eléctrica “Alfredo Desages” (IIIE), Dpto. de Ing. Eléctrica y de Computadoras, CONICET and Universidad Nacional del Sur (UNS), 800 San Andrés Street, Bahía Blanca, Argentina}

\author{Aldo R.\ Fernandes Neto}
\affiliation{Centro Federal de Educação Tecnológica Celso Suckow da Fonseca, Campus Angra dos Reis, Rua do Areal, 522, Pq Mambucaba, Angra dos Reis, RJ, Brazil}

\author{Ben\ Kilminster}
\affiliation{Physik Institut, Universität Zürich, Winterthurerstrasse 190, Zurich, Switzerland}

\author{Patrick\ Lemos}
\author{Katherine\ Maslova}
\author{Irina\ Nasteva}
\author{Ana Carolina\ Oliveira}
\author{Pedro\ Ventura}
\affiliation{Instituto de Física, Universidade Federal do Rio de Janeiro, Av. Athos da Silveira Ramos, 149, Cidade Universitária, Rio de Janeiro, RJ, Brazil}

\author{Martin\ Makler}
\affiliation{International Center for Advanced Studies \& Instituto de Ciencias Físicas, ECyT-UNSAM and CONICET, 25 de Mayo y Francia, Buenos Aires, Argentina}
\affiliation{Centro Brasileiro de Pesquisas Físicas, Rua Dr. Xavier Sigaud, 150, Urca,  Rio de Janeiro, RJ, Brazil}

\author{Franciole\ Marinho}
\affiliation{Instituto Tecnológico de Aeronáutica, São José dos Campos, Brazil}

\author{Jorge\ Molina}
\author{Diego\ Stalder}
\affiliation{Facultad de Ingeniería, Universidad Nacional de Asunción, Campus de la UNA, San Lorenzo, Paraguay}

\author{Laura\ Paulucci}
\affiliation{Universidade Federal do ABC, Avenida dos Estados 5001, Santo André, SP, Brazil}

\author{Dario\ Rodrigues}
\affiliation{Departamento de Física, FCEN, Universidad de Buenos Aires and IFIBA, CONICET, Pabellón I, Ciudad Universitaria, Buenos Aires, Argentina}

\author{Miguel\ Sofo Haro}
\affiliation{Universidad Nacional de Córdoba, CONICET (IFEG) and CNEA (RA0), Córdoba, Argentina}

\collaboration{CONNIE Collaboration}

\date{\today}

\begin{abstract}

The Coherent Neutrino-Nucleus Interaction Experiment (CONNIE) aims to detect the coherent scattering (CE$\nu$NS) of reactor antineutrinos off silicon nuclei using thick fully depleted high-resistivity silicon CCDs\@. 
Two Skipper-CCD sensors with sub-electron readout noise capability were installed at the experiment next to the Angra-2 reactor in 2021, making CONNIE the first experiment to employ Skipper-CCDs for reactor neutrino detection. 
We report on the performance of the Skipper-CCDs, the new data processing, data quality, and event selection for CE$\nu$NS interactions, which enable CONNIE to reach a record low detection threshold of 15\,eV\@.  
The data were collected over 300 days in 2021--2022 and correspond to exposures of 14.9\,g-days with the reactor-on and 3.5\,g-days with the reactor-off. 
The difference between the reactor-on and off event rates shows no excess and yields upper limits for the neutrino interaction rates, comparable with previous CONNIE limits from standard CCDs and higher exposures. 
Searches for new neutrino interactions beyond the Standard Model improve the previous CONNIE limit on a simplified model with light vector mediators. 
A first dark matter (DM) search by diurnal modulation by CONNIE obtains the best limits on the DM-electron scattering cross-section by a surface-level experiment. 
These promising results, obtained using a very small-mass sensor, illustrate the potential of Skipper-CCDs to probe rare neutrino interactions and motivate the plans to increase the detector mass in the near future.

\end{abstract}

\maketitle


\section{Introduction}

Coherent elastic neutrino-nucleus scattering (CE$\nu$NS) is an interaction process in which the neutrino scatters off the nucleus as a whole, benefiting from a coherent enhancement of the cross-section at low neutrino energies. 
Despite being predicted in the Standard Model (SM) over four decades ago~\cite{Freedman:1973yd}, the process was only recently discovered for the first time, thanks to the development of ultra-sensitive detectors and pulsed sources of neutrinos from stopped pions. 
It was first observed by the COHERENT collaboration at the Spallation Neutron Source using a CsI[Na] scintillating crystal detector~\cite{COHERENT:2017ipa} in 2017, and was also detected later using a single-phase liquid argon (LAr) detector~\cite{COHERENT:2020iec} and with a larger CsI[Na] dataset~\cite{COHERENT:2021xmm}.
The observation of the process has opened up a new realm of possibilities for exploring neutrino physics at low energies. 

Since the CE$\nu$NS interaction cross-section is predicted with precision in the SM, its measurement can be used to probe new physics in a number of possible scenarios. 
Nonstandard interactions (NSI) of neutrinos~\cite{Bhupal:2019qno,Papoulias:2019, Lindner:2016wff}, predicted by some extensions of the SM, can be explored at low energies by looking for enhancements of the measured coherent scattering rate due to additional neutral-current processes between quarks and neutrinos. 
Examples of possible searches include models with light scalar~\cite{Dent:2016, Farzan:2018} and vector boson mediators~\cite{Miranda:2020zji, Cadeddu:2020nbr}, or pseudoscalar axionlike particles~\cite{Dent:2019ueq, AristizabalSierra:2020rom}, 
and there is a complementarity in NSI sensitivity between the measurements of stopped-pion and reactor experiments~\cite{Dent:2017mpr}.
Measurements of CE$\nu$NS at short baselines can be used to search for depletion in the flux due to oscillations into sterile neutrinos~\cite{Kosmas:2017zbh, Canas:2017umu, Miranda:2020syh}. 
The process can also probe nonstandard neutrino electromagnetic properties, such as millicharge~\cite{Parada:2019gvy} or charge radii~\cite{Cadeddu:2018dux}. 
Models with a neutrino anomalous magnetic moment, in which the neutrino-nucleus scattering is mediated by a light boson, predict a significant enhancement of the cross-section at low energies and could result in a several orders of magnitude increase in the event rates~\cite{Miranda:2019, Harnik:2012ni}.

The coherent scattering interaction can also be used to measure the weak mixing angle at low energies~\cite{Canas:2018rng, Fernandez-Moroni:2020yyl, Cadeddu:2021ijh}. 
It can provide important information on nuclear structure by measuring the weak form factor, which depends on the neutron density distribution and can be used to measure the neutron radius and the neutron skin~\cite{Hoferichter:2020osn}. 
The process is also relevant to supernova energy transport and stellar collapse~\cite{Freedman:1977xn, Janka:2017vlw} and direct dark matter detection where it is a limiting background~\cite{Monroe:2007xp, Gutlein:2010tq}. 
In terms of potential applications, in recent years the possibility to apply coherent scattering in noninvasive nuclear reactor and nonproliferation monitoring using neutrinos has been investigated~\cite{Bowen:2020unj, Cogswell:2021qlq}.

The recent intense developments in coherent scattering phenomenology and searches for physics beyond the SM have been accompanied closely by rich and varied experimental programmes. 
Detecting CE$\nu$NS is very challenging due to the low neutrino energies involved, $E_{\nu}\lesssim 50$~MeV, and the even lower nuclear recoil energies, typically in the range of tens of eV to few keV. Only about 10\% of the recoil energy contributes to measurable signals such as ionisation, which is reflected in the quenching factors that quantify the ionisation efficiency. 
Therefore the experimental efforts have focused on achieving ever lower energy thresholds and lower background rates, while striving to maximise the detector mass and neutrino flux available. 

On the one hand, in terms of the experiments using stopped-pion neutrino sources, the COHERENT collaboration is taking data and preparing new detectors~\cite{CCM:2021leg}, while new CE$\nu$NS experiments are also being planned for the European Spallation Source~\cite{Baxter:2019mcx}. 
On the other hand, several current and future experiments are using nuclear reactors as a high-flux neutrino source with the aim of detecting CE$\nu$NS.  
The reactor experiments employ a variety of different detection technologies: high-purity germanium crystal detectors in the CONUS~\cite{CONUS:2020skt}, nuGEN~\cite{nGeN:2022uje}, NCC-1701~\cite{Colaresi:2021kus} and TEXONO~\cite{TEXONO:2018nir} experiments, liquid noble gas detectors in RED-100~\cite{Akimov:2022xvr}, scintillating crystals in NEON~\cite{NEON:2022hbk}, cryogenic bolometers and solid-state detectors in MINER~\cite{MINER:2016igy}, NUCLEUS~\cite{NUCLEUS:2019igx} and RICOCHET~\cite{Ricochet:2023yek}, and silicon CCDs (charge coupled devices) in CONNIE and Atucha-II~\cite{atucha2}. 

The Coherent Neutrino-Nucleus Interaction Experiment (CONNIE) seeks to detect the coherent scattering of reactor antineutrinos off silicon nuclei using thick fully depleted high-resistivity silicon CCDs\@. 
The detector is operating in a ground-level laboratory adjacent to the reactor dome of the Angra 2 nuclear reactor near Rio de Janeiro, Brazil, at a position about 30~m from the core of the 3.95 GW thermal power reactor, where the estimated antineutrino flux is $7.8 \times 10^{12}\, \overline{\nu} /\rm{cm}^2/\rm{s}$~\cite{CONNIE2019}.
The experiment had an initial engineering run in 2014 for commissioning and background characterisation~\cite{CONNIE2016}, before installing 14 scientific CCDs in 2016. 
The analysis of 2016--2018 data resulted in an upper limit on the CE$\nu$NS rate of about 40 times the SM expectation for the lowest energy region starting from 75 eV~\cite{CONNIE2019}, which led to competitive limits on simplified NSI models with light scalar and vector mediators~\cite{CONNIE2020}. 
A readout improvement in 2019 data enabled us to lower the detection threshold to 50 eV and obtain new CE$\nu$NS limits~\cite{CONNIE2022}.

In 2021, two Skipper-CCDs~\cite{Tiffenberg:2017aac} were installed in the CONNIE detector with the aim of further decreasing the detection threshold. 
The resulting improvement in sensitivity at low energies is employed to search for CE$\nu$NS, as well as rare processes beyond the SM.
This work reports the first results from the CONNIE experiment with Skipper-CCDs, based on an analysis of the dataset recorded in 2021 and 2022. 
The text is organised as follows. Section~\ref{Sec:Detector} describes the CONNIE experiment detector with Skipper-CCDs and the dataset, Section~\ref{sec:dataprocskpCONNIE} outlines the data processing and calibration procedures, and Section~\ref{sec:Performance} shows the detector performance, event selection and the resulting energy spectra with the reactor on and off. 
These are applied in three 
searches for rare processes in Section~\ref{Sec:searches}: the SM coherent elastic neutrino-nucleus scattering, neutrino NSIs mediated by a light vector particle, and dark-matter diurnal modulations. 
The concluding remarks and prospects are given in Section~\ref{Sec:conclusion}.

\section{The CONNIE detector with Skipper-CCDs}
\label{Sec:Detector}

In order to achieve the sensitivity to low energies required for detecting CE$\nu$NS, it is essential to minimize the readout noise and, consequently, reduce the detection energy threshold of the sensors as much as possible. The recently developed Skipper-CCDs~\cite{Tiffenberg:2017aac}
operate with a nondestructive output readout stage that can sample multiple times the charge in each pixel. 
In this way, the readout noise which is added by the output amplifier is reduced to sub-electron levels and sensitivity to individual electrons is achieved, allowing to count the exact number of electrons in each pixel.  
Skipper-CCD sensors are employed to search for dark matter by the SENSEI~\cite{SENSEI2018}, DAMIC-M~\cite{damicm2020} and Oscura~\cite{Oscura:2023qik} experiments, in the SOAR spectrograph instrument~\cite{Villalpando}, at the Atucha reactor~\cite{atucha2} and in precise studies of silicon properties~\cite{Rodrigues:2020xpt, Botti:2022lkm}. In 2021 two Skipper-CCDs were installed in the CONNIE detector, making it the first experiment to employ these novel sensors for reactor neutrino detection.

\begin{figure*}[tb]
\centering
\includegraphics[height=5.6cm]{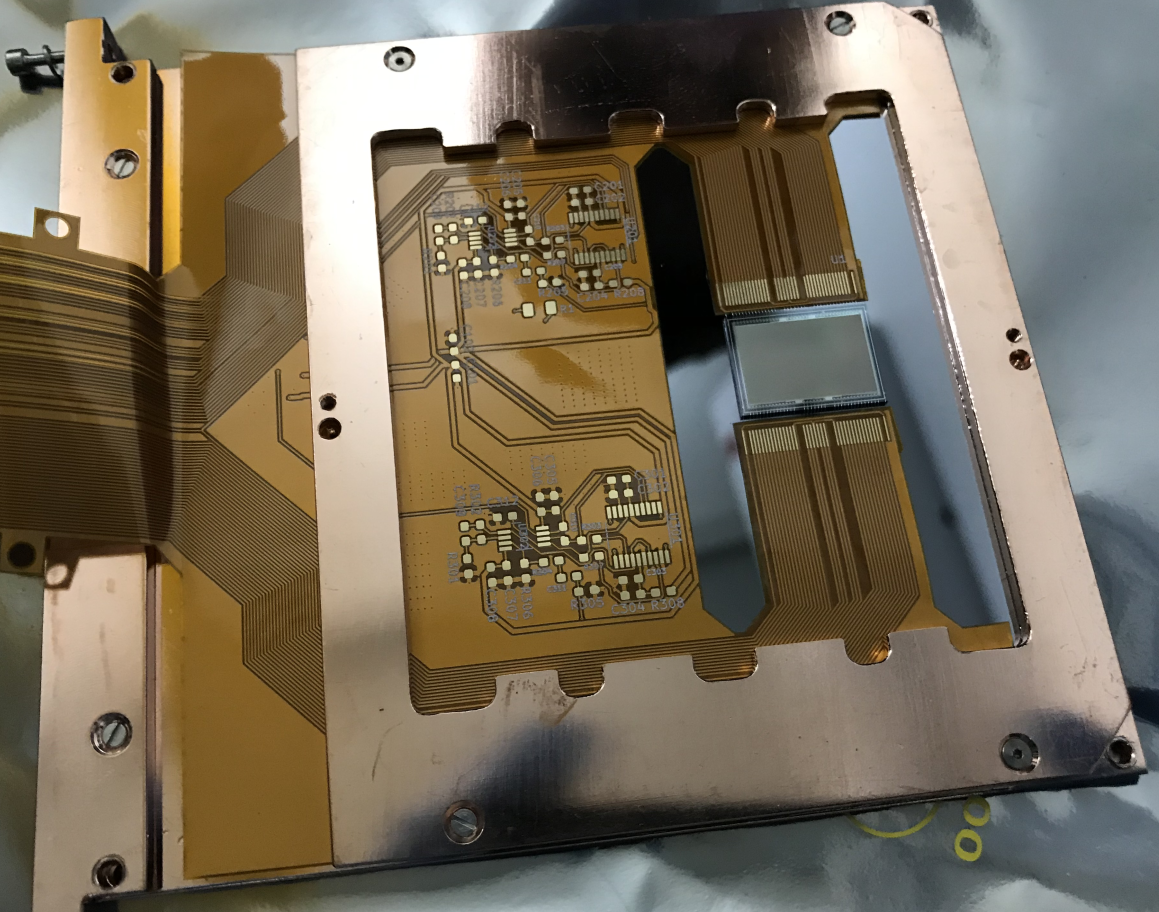}
\includegraphics[height=5.6cm]{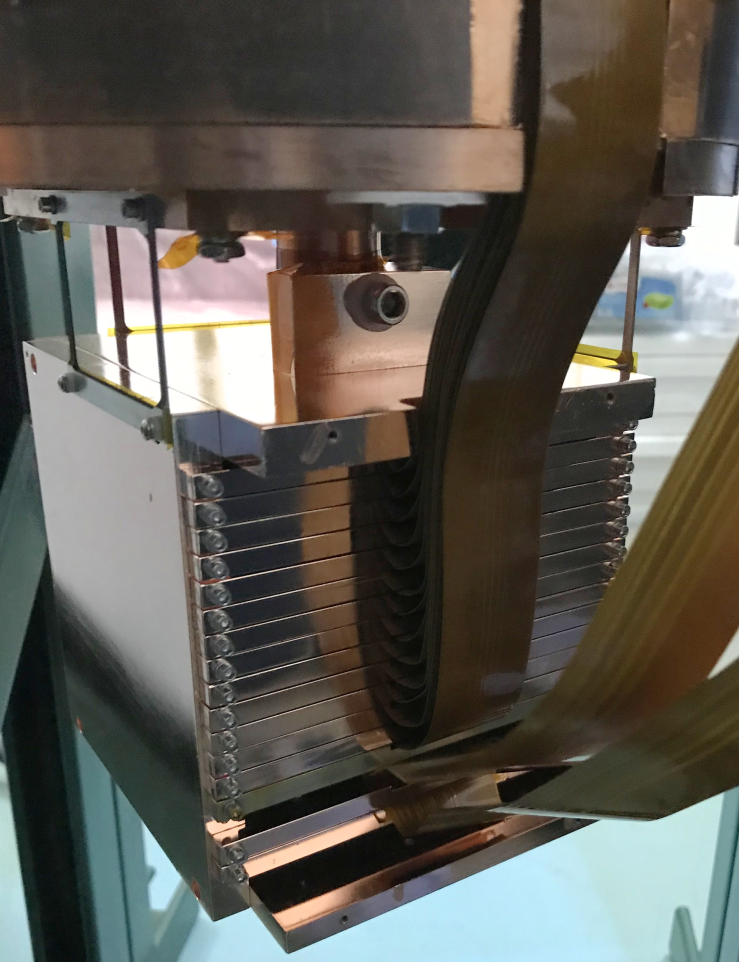}
\includegraphics[height=5.6cm]{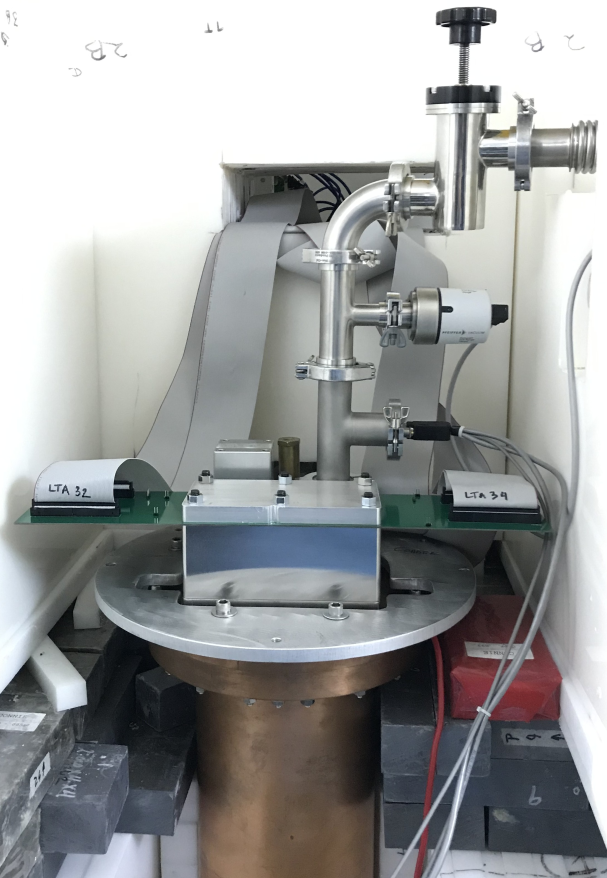}
\caption{CONNIE: (Left) packaged Skipper-CCD, (center) copper cold box with 14 standard CCDs and 2 Skipper-CCDs installed in the bottom positions, and (right) vacuum interface board on top of the copper vacuum vessel.} \label{fig:CONNIEskp}
\end{figure*}

\subsection{Experimental setup and commissioning}

The CONNIE sensors are two identical fully depleted silicon Skipper-CCDs of 675\,$\mu$m thickness, designed at LBNL in collaboration with FNAL during the R\&D phase of the SENSEI experiment~\cite{SENSEI2018} and fabricated at Teledyne DALSA Semiconductor. 
They are made from high-resistivity ($\sim$18\,k$\Omega$-cm) n-type silicon substrate, with a p-type buried channel implant.  
They do not have any backside processing and thus present a thin backside layer with high dopant concentration where partial charge collection occurs, similar to the previous CONNIE CCD sensors~\cite{CONNIE2022}. 
Each sensor is segmented into $1022\times682$ pixels of $15\times15$\,$\mu$m$^2$ size, giving an area of 1.57\,cm$^2$ and a mass of 0.247\,g.

The sensor is connected to a dedicated Kapton flex-cable with microwire bonds, glued to a $7\times7$~cm$^2$ silicon substrate and packaged in a two-piece copper tray, identical to the ones used for the CONNIE standard CCDs, as shown in the left photo of Fig.~\ref{fig:CONNIEskp}. 
The two Skipper-CCD packages were installed in the lowest slots of the copper box that holds the sensors in the detector, while the 14 standard CCDs from the previous CONNIE setup remained in their original positions (Fig.~\ref{fig:CONNIEskp} center). 
The Skipper-CCD flex cables connect through new dedicated second-stage flex circuits with signal preamplification to the
vacuum side of a newly designed vacuum-interface board (Fig.~\ref{fig:CONNIEskp} right), which transfers the signals from the two Skipper-CCDs and two standard CCDs\@. 
From the air-side of the interface, the signals are taken to four Low-Threshold-Acquisition (LTA) readout boards~\cite{Cancelo2021} using long flat cables. 

The rest of the detector setup is unchanged with respect to the previous operations~\cite{CONNIE2019,CONNIE2022}. 
The sensors operate in a vacuum of 10$^{-7}$~torr and are cooled to below 100\,K (measured at the cold finger) to minimise thermally generated dark current without loss of charge-transfer efficiency. The detector is surrounded by passive shielding, including 15 cm of lead to absorb photons and two 30-cm layers of high-density polyethylene to shield against cosmogenic neutrons.
After the installation, the experiment took data for commissioning from mid-July to early November 2021, when the full passive shielding was assembled. 
Since then, the data acquisition has been continuous until the end of 2023, with short interruptions due to technical issues, on-site interventions or power cuts.

\subsection{Data acquisition}

The best operational configuration of the 2 Skipper-CCDs in terms of the quality of the acquired images was achieved by measuring half of each sensor using one of the 4 output amplifiers located in its corners.
The data acquisition cycle consists of two phases: cleaning and readout. 
During cleaning, a procedure to reduce surface dark current is performed, in which the sensor surface voltages are reversed~\cite{Holland:2003kiw}.
Then, after reverting the voltages to their default values, the charge in the CCD active area is clocked and discarded for 10 minutes (equivalent to reading out the entire sensor about 15 times with only one sample per pixel), ensuring that the readout starts with a ``clean" CCD\@. 
The readout is done sequentially by alternating the phasing of the vertical clocks, so that the charge in the row of pixels adjacent to the serial register is first moved into it.
Subsequently, the entire row of pixels is retrieved as the charge is sequentially shifted through the amplifier in the serial register, guided by another set of three voltages known as the horizontal clocks.

Both sensors are read out simultaneously 
with \mbox{$N_{\rm skp}=400$} samples per pixel, taking approximately 2~hours to acquire a raw image and resulting in a nonuniform exposure in different pixels. 
The data-taking periods are divided into runs, typically of $\sim$500 images  acquired under stable detector configuration and operating conditions. New runs are also started after interruptions, to ensure clear segmentation and traceability of the data.

The current dataset contains 300 days of data, collected between November 2021 and December 2022, from which 243 days were taken with the reactor on. 
There were two periods when the reactor was off: from June 12 to July 25, 2022 and from November 10 to 29, 2022, during which a total of 57 days of reactor-off data were collected.

\section{Data processing} \label{sec:dataprocskpCONNIE}

The raw data are two-dimensional FITS images~\cite{fitsformat} in which each group of $N_{\rm skp}$ consecutive pixels in a row corresponds to $N_{\rm skp}$  measurements of the charge of a single pixel. 
The first processing step 
constructs a two-dimensional image assigning to each pixel the average of the $N_{\rm skp}$ charge measurements, in analog-to-digital units (ADU)\@. 
The horizontal baseline is then computed for each row of the processed image as the median of the pixels in the overscan (unexposed) region, while the vertical baseline is obtained for each column as the median of the first 100 pixels in the column.
The horizontal and vertical baselines are subtracted from each row and column, respectively.
The images subsequently undergo energy and size-to-depth calibrations, as well as masking of defects and spurious charges, before the events  corresponding to energy depositions in the sensors are extracted.

\begin{figure}[tb]
 \centering
  \includegraphics[width=\linewidth]{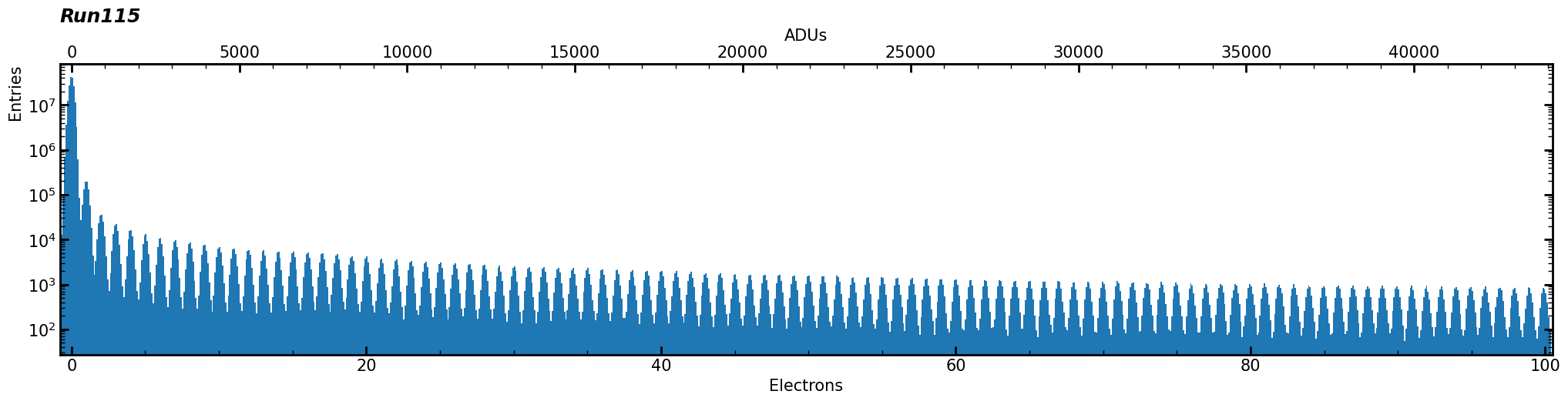}
 \caption{Distribution of charge up to one hundred electron peaks in the data from one run.}  \label{fig:electronpeaks}
\end{figure}

\subsection{Energy calibration}

Due to the Skipper-CCD readout noise achieved with $N_{\rm skp}=400$~samples/pix reaching a fraction of an electron (as detailed in Section~\ref{sec:data_quality}), the consequent charge distribution of the images is a sequence of Gaussian peaks with the means centered on multiples of the electron charge, as shown in Fig.~\ref{fig:electronpeaks}. 
This individual electron resolution of the sensors \cite{Tiffenberg:2017aac} is used to calibrate the image from ADU to number of electrons.

In order to establish a precise calibration for all the images in a run, the distributions of charges per pixel are obtained for all images in a run, resulting in samples of sufficient size. 
This is shown in Fig.~\ref{fig:electronpeaks} for the first 100 peaks of one sample, which are then fitted with independent Gaussian functions. 
A linear fit is performed to the obtained Gaussian means to derive the global gain from the slope and the baseline residual correction from the intercept.  The nonlinearity obtained with this method in the energy range of interest is less than 1\%, in good agreement with previous works~\cite{Rodrigues:2020xpt}.

The gain of each image is also monitored throughout the data taking by performing a simplified fit to the first two Gaussian peaks in the charge distribution, in order to test the gain stability. 
Images with outlier values of the gain are excluded from further analysis.

\begin{figure}[tb]
\centering
\includegraphics[width=0.4\textwidth]{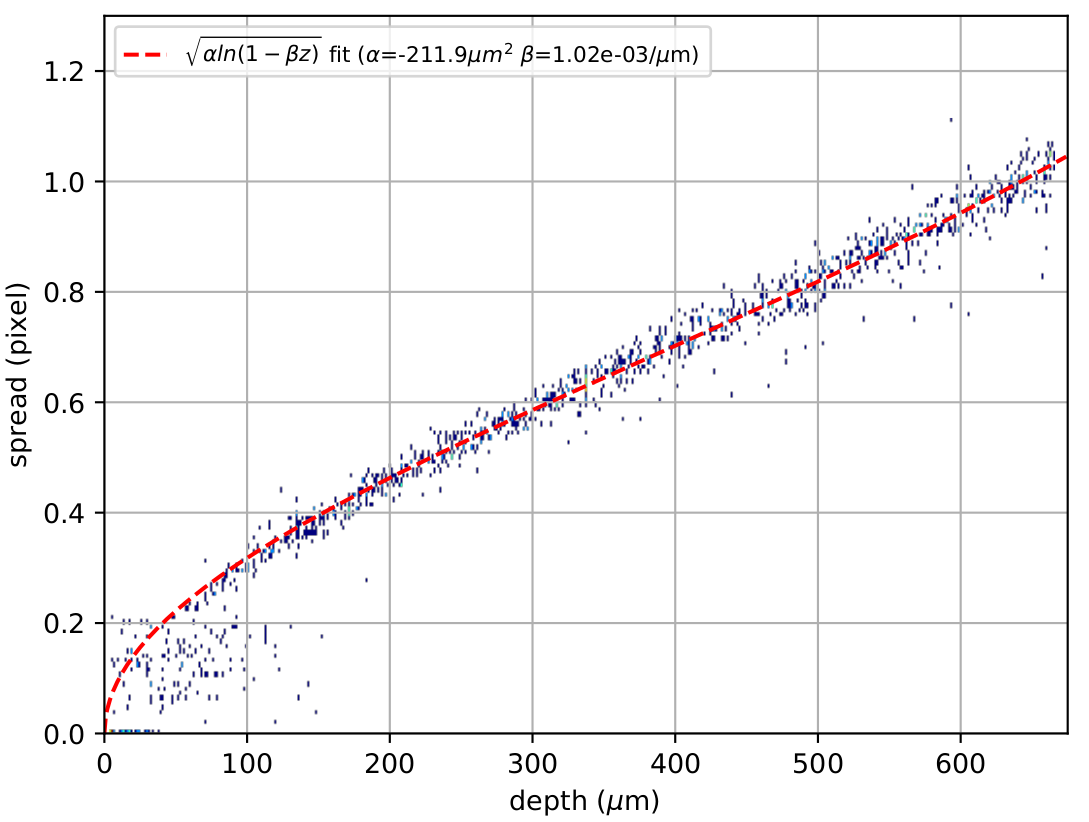}
\caption{Size-to-depth calibration curve from a sample of 45 muons in one Skipper-CCD.} \label{fig:sizedepthcal}
\end{figure}

\subsection{Size-to-depth calibration}

When a pointlike energy deposition within the sensor volume produces a number of charge carriers (holes), the electric field present in the bulk causes them to drift towards the potential wells at the pixel top surface of the CCD where they are collected. 
As they drift, the holes diffuse laterally in the plane perpendicular to the drift direction ($z$). 
As shown in Ref.~\cite{SofoHaro:2019ewy}, the lateral spread is Gaussian with a variance that depends on the time the holes diffuse before being collected, which is proportional to the depth of the ionisation event. Holes produced close to the CCD back side have more time to diffuse before being collected and, therefore, they spread more than the holes produced close to the collection wells.

The size-to-depth calibration curve is determined from muon tracks using the same procedure as in standard CCDs~\cite{CONNIE2022}. 
Cosmogenic muons pass through the CCD leaving a straight track which is narrow at the entry point at the top of the CCD and widens as it progresses towards the back surface, where it exits. 
A sample of muon tracks perpendicular to the horizontal register is used to determine the lateral spread as a function of depth, ensuring an unambiguous depth assignment along the track and resulting in a clean and ubiased size-depth calibration, despite the reduced sample size. 
Since the muon trajectory is  straight, a depth can be assigned to each muon slice and therefore to each value of the lateral spread, hence composing the calibration curve. 

The behaviour of the size-to-depth relation from a sample of 45 muons in one sensor and one typical run is plotted in Fig.~\ref{fig:sizedepthcal}. Each point is a measurement of the size and depth of one slice of a muon track. The fitted curve $f(z)=\sqrt{\alpha\ln(1-\beta z)}$ with parameters \mbox{$\alpha = -212~\mu{\rm m}^2$}, $\beta = 1.02\times10^{-3}\mu{\rm m}^{-1}$, shown by the red dashed curve, gives the spread in~$\mu$m, with variations of up to 4\% between sensors and runs. 
Deviations of the fit at small depths reflect the limited reconstruction resolution at small event widths and do not bias the fit parameters in the remaining fit range.
The event size depends on the energy deposit per pixel, and muon-based calibration curves slightly overestimate the event size for low-energy interactions due to the more spatially spread muon signals caused by charge repulsion, whereas neutrino-induced deposits undergo diffusion only; this difference is below 7\% in the low-diffusion regime and decreases with increasing diffusion, and is therefore neglected~\cite{SofoHaro:2019ewy, SofoHaro2017}.

\subsection{Masking}  
\label{sec:Masking}

Due to the sub-electron noise, an order of magnitude lower than standard CCDs, the Skipper-CCD sensors have a high image resolution, making it possible to investigate artifacts of charge smearing and defective pixels that can mimic low-energy events from diffusion-limited hits and become a background source.
To eliminate these features, a masking procedure to flag these background events in the images was developed using reactor-off data and empty images formed by reading out serial registers without movement of pixel charges, and was applied prior to the event extraction.

The features known as serial-register events (SRE) manifest in the images as charge deposits along a single row, originating from charge diffusion in the inactive silicon region that extends to the serial register. They frequently occur in clusters of adjacent pixels and occasionally in more dispersed arrangements along a single row. The latter may include isolated pixels reconstructed independently from the group of pixels associated with a single SRE, leading to potential misidentification as a diffusion-limited hit mimicking a neutrino event.

Consequently, for each image, an individual mask is generated in which entire rows with SRE candidates are marked for exclusion. 
The selection requires the presence of at least two pixels in the same row, each with a minimum charge of $1.5\,{\rm e^-}$, not separated by more than two pixels, and with no adjacent pixels above and below with a charge greater than $1.5\,{\rm e^-}$. 
This threshold value was chosen to exclude single-electron events, such as spurious charges or dark current~\cite{SENSEI:2021hcn}. 
The SRE identification algorithm was tested on images containing only serial-register events. The resulting masks correctly identified more than 99\% of the SREs in these images.

In addition to the SRE, another common background source in CCDs consists of  hot pixels that produce bright columns or rows, or appear as individual hot pixels. 
These are identified by searching the image regions without particle tracks for rows or columns with more than 12 unmasked pixels with a charge greater than $0.6\,{\rm e^-}$. The value of 12 was chosen based on the observed distribution of pixels exceeding this threshold.
To exclude hot features that appear  recurrently throughout a run, a single mask per run is built containing the flagged columns and rows that were found hot in more than 5$\%$ of the images, since this fraction marks the appearance of clear outliers in the frequency distribution of such events.
Similarly, individual pixels that have charges above $2.56\,{\rm e^-}$ (16 times the readout noise) in more than 10\% of the images are flagged as bad pixels.
Merging the two hot pixel masks and the SRE individual masks results in an individual global mask for each  image in a run. 

The average fraction of pixels that were flagged in the global masks corresponds to 8\% of the  Skipper-CCD areas (6\% for one sensor and 10\% for the other). 
The impact of the masking, concentrated on a few hot columns and rows, is a reduction by almost two orders of magnitude in the total measured charge.

\subsection{Event extraction}
\label{sec:eventextraction}

Once the images are calibrated and masked, an event extraction procedure is applied to identify the pixel clusters associated to energy depositions and to create an event catalog for each image. 
The event extraction algorithm is the same as the one described in Ref. \cite{CONNIE2019}, but uses  a charge threshold of 1.6$\,{\rm e^-}$ ($10\sigma$ above the noise level) for the seed pixel and of $0.64\,\rm e^-$ ($4\sigma$) for its adjacent pixels. 
These values were chosen in order to reconstruct the total event charge.
A cluster is formed by grouping the seed pixel and its neighbouring pixels (including diagonal connections) until no additional adjacent pixels above threshold are found. A catalog containing the properties of all reconstructed clusters is saved for each image.

\begin{figure*}[tb]
\centering
\includegraphics[width=0.75\textwidth]{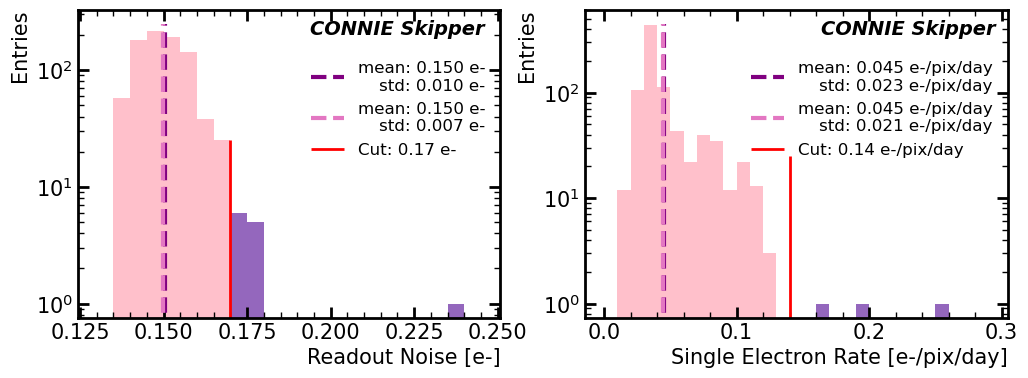}
\caption{(Left) Readout noise and (right) single-electron-rate distributions for reactor-off images. The purple histograms represent the whole sample, and the pink histograms the data retained after the cut.} \label{fig:perfromance}
\end{figure*}

Additional event size and shape variables computed in the extraction algorithm are the event widths $\sigma_x$ and $\sigma_y$, obtained from a two-dimensional Gaussian fit to the cluster charge spatial distribution. 
The coordinates of the barycenter position of the event in the image are also calculated. 
The subsequent data selection is performed on the image catalogs containing the event variables.

\section{Selection and energy spectrum}
\label{sec:Performance}

\subsection{Data quality criteria}
\label{sec:data_quality}

To establish a set of data quality criteria, the performance of three parameters was analysed: the readout noise, the single-electron event rate and the number of flagged pixels in the global mask. 
The analysis was performed using only reactor-off data in order to avoid experimenter's bias. 

The readout noise is the fluctuation added by the output amplifiers, while the single-electron events~\cite{SENSEI:2021hcn} correspond to charges caused by a combination of different processes such as spontaneous thermal emission or dark current, Cherenkov and recombination photons~\cite{Du:2020ldo} or clock-induced charge or spurious charge~\cite{SENSEI:2021hcn}.
Both the noise and the rate of single-electron events of each image are calculated during the data processing 
and continuously monitored, together with the gain. 

The procedure to compute the noise and single-electron rate is analogous to the one in the analysis of standard CCDs~\cite{CONNIE2022}, and calculates the two parameters independently from the overscan and active regions, respectively. 
In addition to applying the SRE mask, all pixels with more than 5\,e$^-$ and their 10 adjacent pixels in all directions are excluded from the active region in the calculation, leaving contributions from mostly pixels with 0 and 1 electron.

Figure~\ref{fig:perfromance} shows the readout noise and single-electron-event rate distributions of one Skipper-CCD\@.
The mean readout noise of the reactor-off dataset is measured to be ($0.150\pm 0.007$)\,e$^-$. 
In order to exclude the outlier noisy images, only those with noise below 0.17\,e$^-$ are retained in the subsequent analysis. 
The single-electron rate is obtained as ($0.045\pm 0.021$)\,e$^-$/pix/day. 
Since single-electron events can interfere significantly in the background rates, images are required to have single-electron rate of less than 0.14\,e$^-$/pix/day.
Lastly, to ensure a good spatial coverage, images with more than 25\% of masked pixels in the active region were excluded from the analysis.
The effect of the data quality cuts is a reduction of 2.9\% (5.6\%) in the reactor-off (on) data sample.

\begin{figure*}[tb]
\centering
\includegraphics[height=5.4cm]{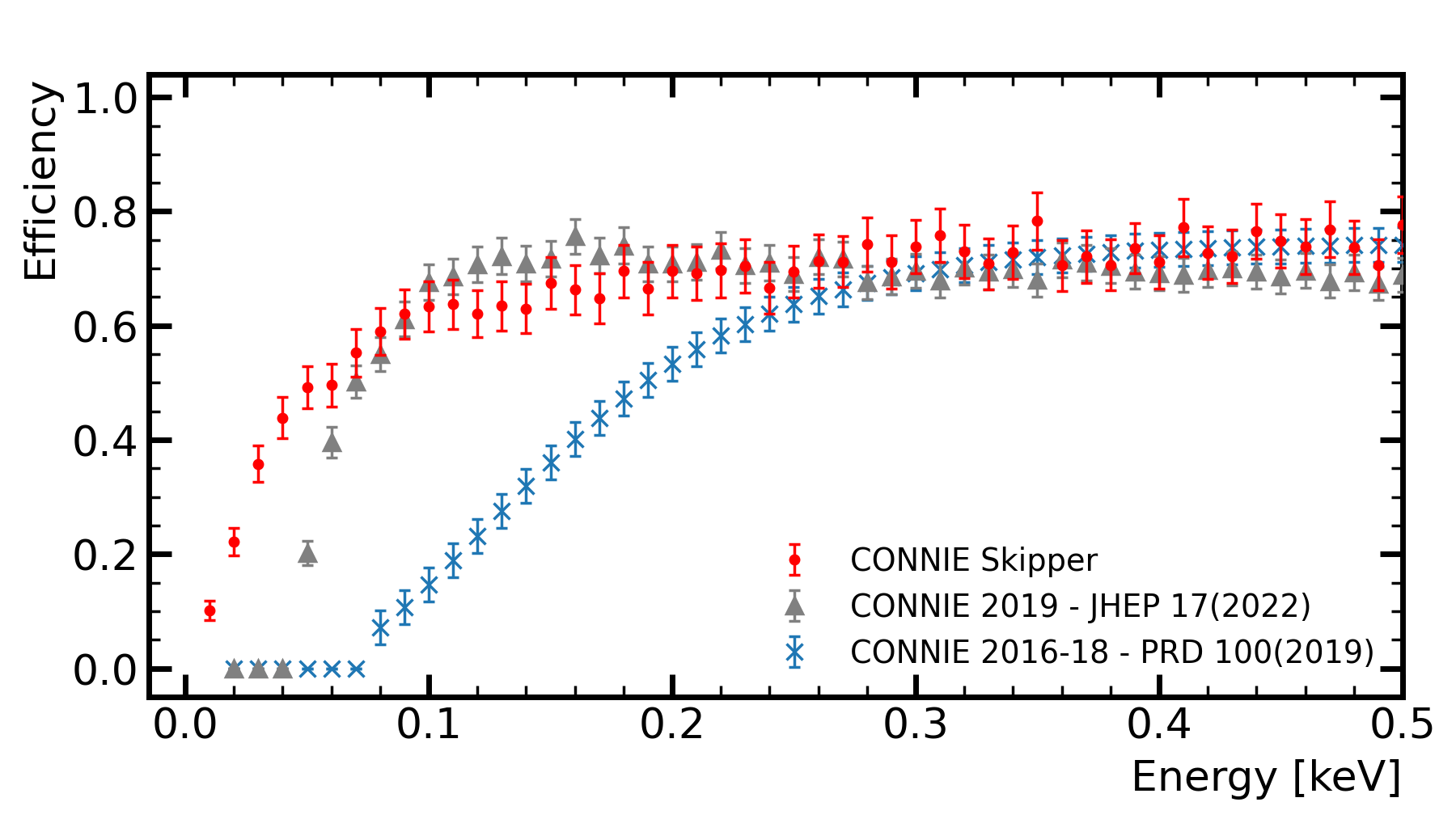}\hspace{.1cm} 
\includegraphics[height=5.4cm]{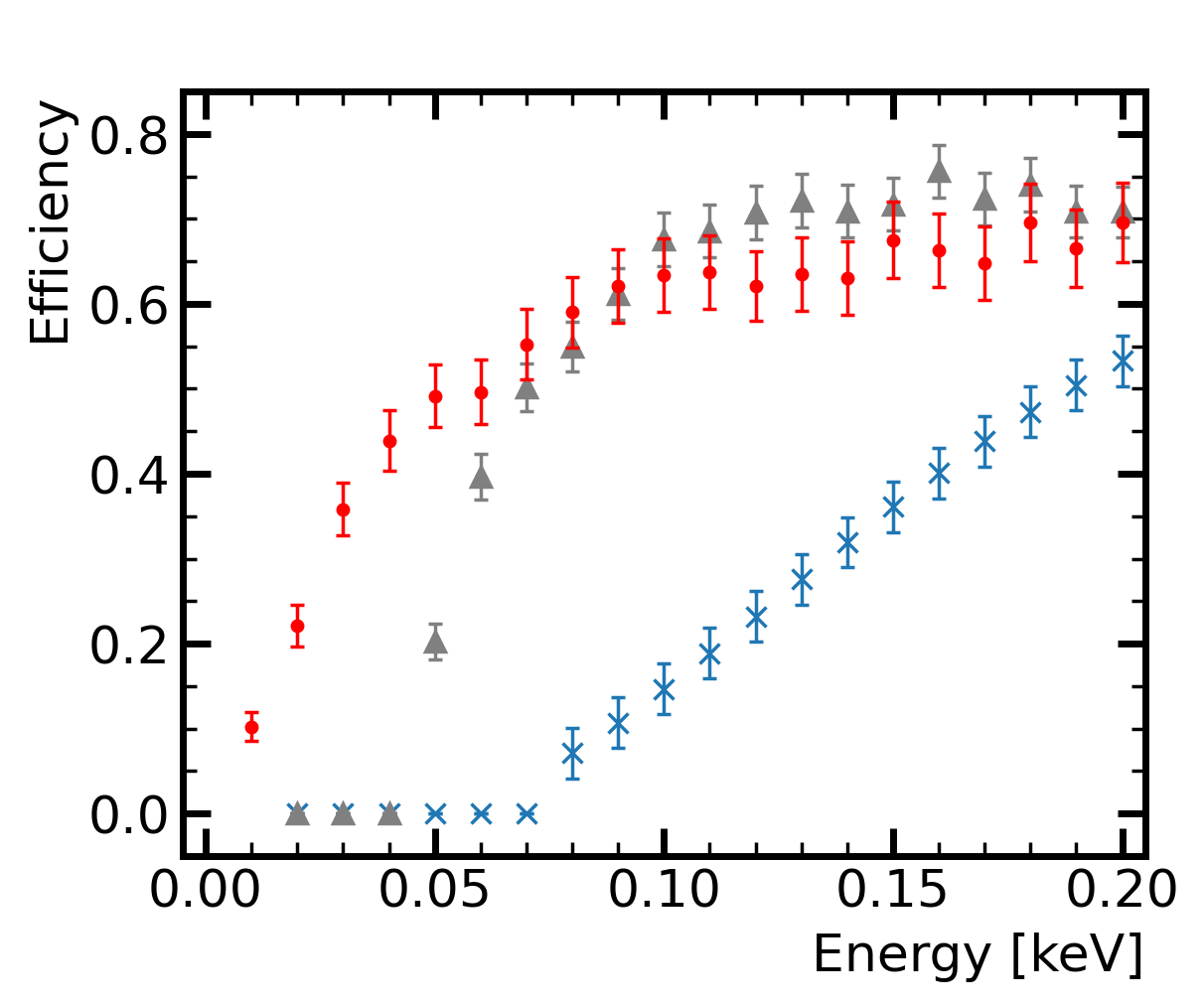}
\caption{CONNIE Skipper-CCDs detection efficiency (red), accounting for the event extraction acceptance and the selection cuts. The CONNIE standard CCD efficiencies from 2016-18~\cite{CONNIE2019} (blue) and 2019~\cite{CONNIE2022} (grey) are shown for reference. (Left) Efficiency up to energies of 0.5\,keV and (right) a zoom into the low-energy region.}
\label{fig:effCONNIEskp}
\end{figure*}

\subsection{Event selection and exposure}

After the data quality requirements, event selection criteria are applied similarly to the previous analysis~\cite{CONNIE2022}, based on the event positions within the active region and their shapes, in order to search for neutrino interactions in the sensor volume. 
First, the image borders of 10 pixels in the active region are excluded, in order to avoid edge effects in the morphology of reconstructed events~\cite{janesick2001scientific}. 
This results in a reduction of around 8\% in the effective area of the sensor and is reflected in the calculation of the total exposure. 

Then a selection is applied based on the size of the events. 
The events are required to have widths $\sigma_x$ and $\sigma_y$ smaller than 0.95 pixels to exclude charge deposits that are very spread or come from the back CCD layer which has only partial charge collection and can produce fake low-energy events~\cite{Fernandez-Moroni:2020abn}. 
In addition, a lower limit of 0.2 pixels is set on the widths, considering the sensitivity of the reconstruction algorithm. These requirements result in rejecting about half of all previously selected events.

Since the Skipper-CCDs do not have an exposure time prior to the readout stage, a single pixel in the sensor is only exposed during the sequential readout of other pixels before it in the readout line. 
This is taken into account to calculate the effective exposure of each image.
A time map is built assigning to the $n$-th pixel an exposure time $t_{\rm exp} = t_{\rm ro}(n-1) / (N_{\rm pix}-1)$, where $t_{\rm ro}$ is the total image readout time and $N_{\rm pix}$ is the total number of pixels in the image. 
After removing all the regions excluded in the geometrical border cut and the individual global mask, the image exposure time is calculated as the average exposure time of the remaining pixels in the time map.

\subsection{Detection efficiency}
\label{sec:Efficiency}

The detection efficiency determination uses the same methodology as in previous analyses~\cite{CONNIE2019, CONNIE2022} by simulating 40 neutrino-like events per image from the reactor-off period, with a uniform probability in the active volume and a uniform energy distribution between 5 and 2005\,eV, using the depth-size calibration. 
Comparing the energy and three-dimensional position of the simulated events before and after the image processing and event extraction, the total extraction acceptance was found to be 98.7\%. 
The remaining small fraction of simulated events was flagged by the masking routine due to their similarity to serial-register events, Section \ref{sec:Masking}.

The overall detection efficiency, shown in Fig.~\ref{fig:effCONNIEskp}, accounts for the event extraction acceptance and the selection procedure.
Compared to the previous runs with standard CCDs, the new sensors extend their efficient operation to lower energies and reach full efficiency around 100\,eV\@, representing a significant improvement.
Based on this, the energy threshold for rare event searches was reduced to 15\,eV\@. This value was chosen based on the limitation of the current understanding of the background at very low energies, where artificial events are not yet fully characterised or efficiently rejected, leading to an increase in the observed background levels. With this threshold, the experiment achieves the lowest-energy sensitivity among all current CE$\nu$NS searches, improving over the previous best limit of 50\,eV also obtained by CONNIE~\cite{CONNIE2022}.

\begin{figure}[tb]
\centering
\includegraphics[width=\linewidth]{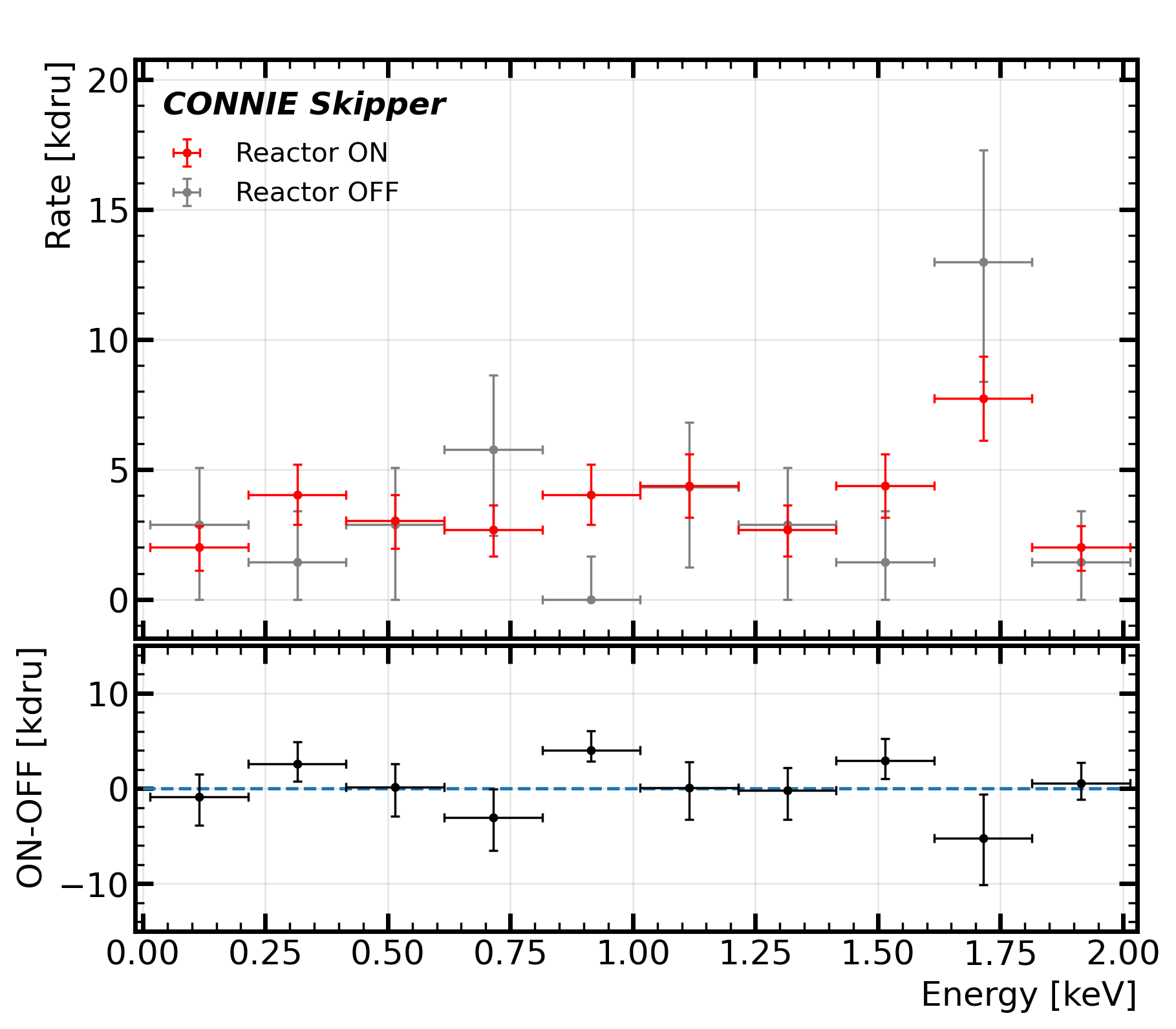}
\caption{(Top) CONNIE Skipper-CCD reactor-on and reactor-off spectra and (bottom) their difference.}
    \label{fig:SpectrumCONNIEskp}
\end{figure}

\subsection{Energy spectrum}

All the processing, data quality and event selection criteria were fixed before unblinding the reactor-on data. 
The total effective exposures of the data samples after all selection requirements were determined to be 3.5\,g-days with the reactor off and 14.9\,g-days with the reactor on. 

The event-rate distributions of the energy spectra for the selected events over the two periods, accounting for their effective exposure, are shown in Fig.~\ref{fig:SpectrumCONNIEskp} (details in Appendix~\ref{AppendixRates}). 
The mean event rates are comparable at approximately 4\,kdru, and are consistent with the previous analyses~\cite{CONNIE2019, CONNIE2016, CONNIE2022}. 

The difference between the reactor-on and off energy spectra is also shown, demonstrating no significant excess. The p-values for the null (background only) hypothesis are estimated using the statistical prescription of Ref.~\cite{Cousins:2007yta} for the first three bins to be equal to 0.82, 0.26 and 0.65, respectively. These values do not allow to reject the null hypothesis at the 90\% confidence level. 
Nevertheless, due to the low energy threshold and the low single-electron-event rates, the current dataset can be employed to search for rare processes in and beyond the SM.

\section{Searches for rare processes}
\label{Sec:searches}

\begin{table*}[tb]
\renewcommand{\arraystretch}{1.5}
\centering
\begin{tabular}{ c | c| c | c | c }
Measured Energy   & Sarkis (2023) rate & Chavarria rate  &  Observed $95\%$ CL & Expected 95\% CL \\ 
$[\rm{keV_{ee}}]$ &  [$\rm{kg^{-1}d^{-1}keV_{ee}^{-1}}$]& [$\rm{kg^{-1}d^{-1}keV_{ee}^{-1}}$] &  [$\rm{kg^{-1}d^{-1}keV_{ee}^{-1}}$] &  [$\rm{kg^{-1}d^{-1}keV_{ee}^{-1}}$]\\ 
\hline
$0.015-0.215$ &  $29.3\,^{+4.6}_{-4.7}$ & $\,17.7\pm 3.3$ & 2.24$ \times 10 ^3$ & 3.18 $ \times 10 ^3$ \\ 
$0.215-0.415$ & $2.7\,^{+1.3}_{-1.2}$ & $\,2.20\pm 0.21$ & 7.36$\times 10 ^3$ & 4.77 $ \times 10 ^3$\\ 
$0.415-0.615$ & $0.43\,^{+0.41}_{-0.39}$ & $0.36\pm 0.04$ & 3.41$\times 10 ^3$ &  3.31 $ \times 10 ^3$ \\ 
\end{tabular}   
\caption{Expected CE$\nu$NS event rates for different ionisation efficiency models considering the efficiency and resolution for Skipper-CCDs. In the last column the observed and the expected limit at 95\% CL are tabulated. }\label{Rates}
\end{table*}

\subsection{Search for coherent elastic neutrino-nucleus scattering}

In order to obtain the expected neutrino rates at the detector, the SM neutrino coherent interaction rates are corrected for the effects of selection efficiency and resolution, obtained from simulation, as well as for the quenching factor that represents the fraction of nuclear recoil energy that creates ionisation. Here a similar procedure is used as the one in Ref.~\cite{ChavarriaQF}.

The predicted antineutrino flux at the detector is calculated following the summation method described in Refs.~\cite{NufluxSumationMeth,ReactorNuSumationPerise,ReactorNuSumationPerise2,MILLS20202130}, updated with improved antineutrino spectra for the fissile isotopes $^{235}\rm{U}$, $^{238}\rm{U}$, $^{239}\rm{Pu}$ and $^{241}\rm{Pu}$. 
The contribution from activation (neutron capture) of structural and fuel elements relevant for energies below 1.27 MeV was also updated following \cite{FluxNuT}. The detection threshold of 15\,$\rm{eV_{ee}}$ corresponds to a minimum observable neutrino energy of $\sim$0.44~MeV, according to the Sarkis {\it et al.} quenching factor model~\cite{Sarkis2023}. Above this energy, the updated flux model and the one used previously \cite{ChavarriaQF} agree to within 3\% and are thus considered equivalent in this analysis 
(details can be found in Appendix~\ref{AppendixRates}).

The detection efficiency discussed in Section~\ref{sec:Efficiency} is used as a multiplicative correction to the expected event rate as a function of the measured ionisation energy, $E_M$.
For the quenching factor, $f_n(E_{\rm nr}) = E_I/E_{\rm nr}$, with $E_I$ the ionisation energy, the updated model of Sarkis {\it et al.}~\cite{Sarkis2023} is applied, which allows to calculate the ionisation efficiency for nuclear recoil energies in silicon down to 0.05\,$\rm{keV_{nr}}$. 
This model, based on the original ideas by Lindhard \cite{lindhard}, incorporates improved descriptions of the electronic stopping, interatomic potential and electronic binding at sub-keV energies, covering the Skipper-CCD low-energy range ($E_{\rm{nr}}>0.240 \,\rm{keV_{nr}}$) corresponding to  $E_M>0.015$\,keV$_{\rm ee}$.
The energy resolution $R(E_I,E_M)$ is modeled as a Gaussian with variance $\sigma^2=\sigma^2_0 +F_eE_{eh}E_I$, where $\sigma_0=3.04$ $\rm{eV_{ee}}$ is the readout noise, $F_e=0.119$ is the electronic Fano factor, and $E_{eh}=3.752$\,$\rm{eV_{ee}}$ is the mean energy to create an electron-hole pair in silicon~\cite{Rodrigues:2023fwi}. 

The resulting predicted rates for the CE$\nu$NS  spectrum for Skipper-CCDs are shown in Table \ref{Rates}, for the same binning as used in the data.
A comparison with the Chavarria quenching factor~\cite{Chavarria:2016xsi}, valid for $E_{\rm{nr}}>0.30 \,\rm{keV_{nr}}$, and out of range for Skipper-CCD data near the threshold, is also given for reference. 
The uncertainties in the expected rates take into account systematic variations due to the quenching factor \cite{Sarkis2023,Chavarria:2016xsi}, efficiency (see Appendix~\ref{Appendix1} for details) and reactor budget \cite{PerisePWRErr,ReactorNuSumationPerise2}.

Based on the difference between the measured reactor-on and off event rates in Fig.~\ref{fig:SpectrumCONNIEskp}, a 95\% confidence-level (CL) upper limit on the observed CE$\nu$NS rate is established~\cite{Barlow:2003xcj, Barlow:2004wg}, shown in Table~\ref{Rates}. 
The observed limit in the lowest-energy bin, where the expected rates reach the highest values, corresponds to 76 times the SM prediction using the Sarkis quenching factor~\cite{Sarkis2023}.
The expected limits are also calculated and shown in the last column of the table.
The observed limit  obtained with Skipper-CCDs and the previous result using standard CCDs~\cite{CONNIE2022} are of similar size. 
This can be explained by the fact that, on the one hand, the sensitivity of the current limit is restricted by the large rate uncertainties due to the low exposures, while on the other hand it benefits from the significant improvement in detection efficiency at low energies. 
This enhanced sensitivity motivates the need to increase the sensor mass in order to harness the full potential of Skippers-CCDs in the search for CE$\nu$NS with reactor neutrinos.

\subsection{Search for new light vector mediator with CE$\nu$NS detection channel}

\begin{figure*}[tb]
\centering
\includegraphics[width=0.49\textwidth]{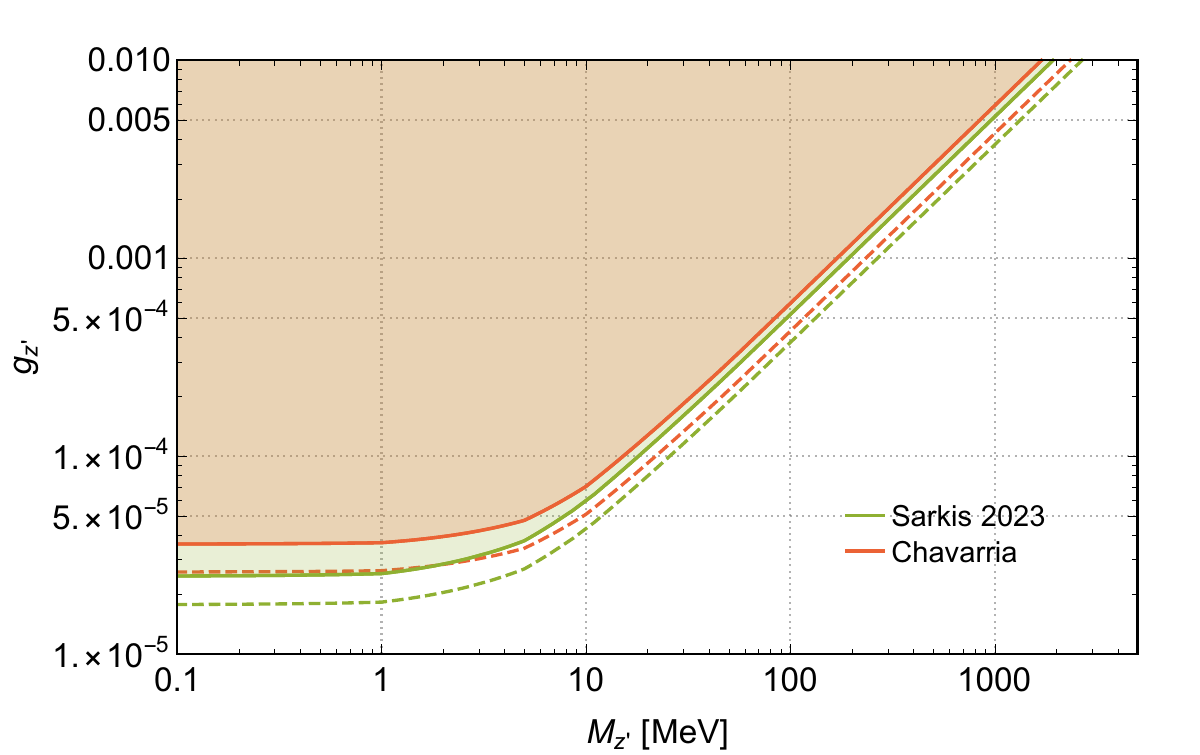}\hfill
\includegraphics[width=0.49\textwidth]{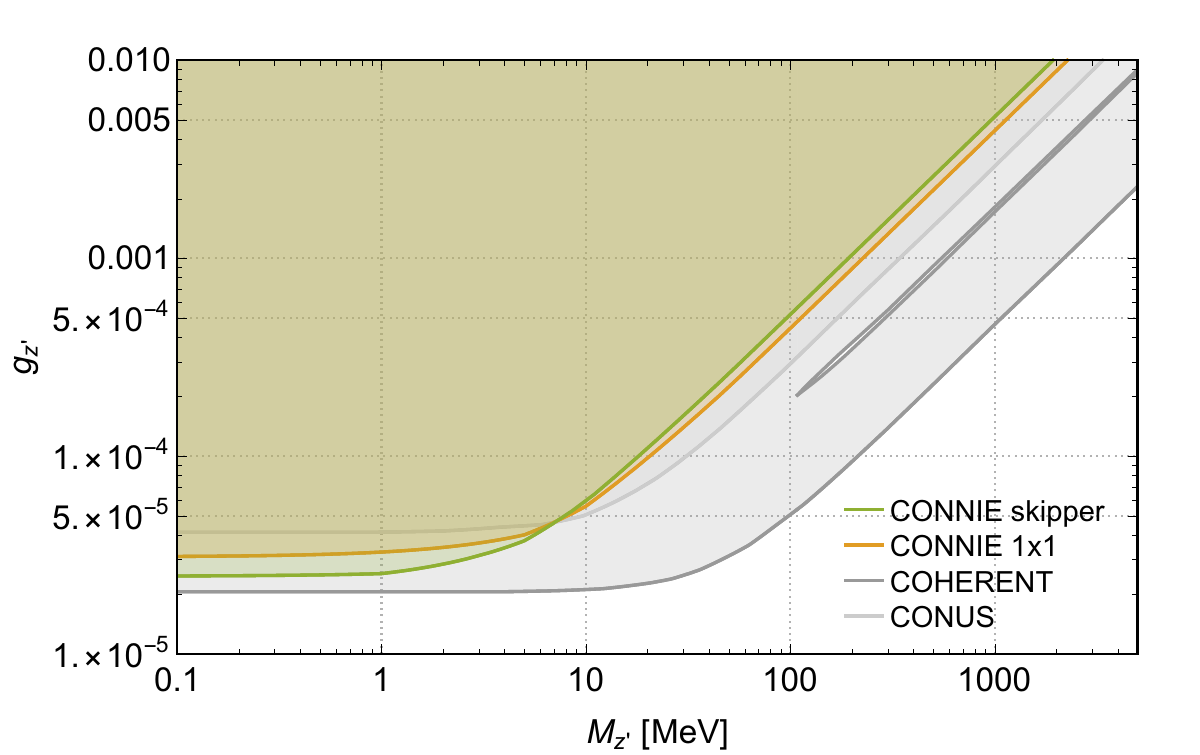}
\caption{(Left) CONNIE Skipper-CCD exclusion limits (solid lines) at 95\% CL from the CE$\nu$NS detection channel for a light vector mediator $Z^{\prime}$, assuming the simplified universal model~\cite{Cerdeno_2016}, and considering quenching factors from Sarkis~\cite{Sarkis2023} (green) and Chavarria~\cite{Chavarria:2016xsi} (red); and projections of these limits (dashed) assuming a zero difference between the rates and a five times smaller statistical error. 
(Right) Comparison between CONNIE exclusion limits at 95\% CL from Skipper-CCDs considering the quenching factor from Sarkis~\cite{Sarkis2023} (green) and the standard-CCD~\cite{CONNIE2020} (red). For comparison, limits at the 90\% CL from COHERENT (CsI+Ar)~\cite{Corona:2022} and CONUS ($k=0.16$) data~\cite{CONUS:2022} are also shown.} \label{fig:LVlim}
\end{figure*}

Theories beyond the SM that include light mediators are highly motivated by the experimental results in various fields of particle physics~\cite{Fayet2007, Kaplan2009, Langacker_2009, Giunti2015}, including dark matter and neutrino physics. New mediators can be probed under the framework of the so-called ``simplified models'' which, accounting only for representative new particles and interactions, allow to characterise new physics with a small number of parameters~\cite{Morgante_2018}. Reactor neutrino experiments, sensitive to low-energy interactions, have demonstrated their competitiveness in constraining new light mediators in the mass region below $\sim$10~MeV~\cite{CONNIE2020, CONUS:2022}.

Here, the CONNIE results from the Skipper-CCD run are employed to constrain the parameter space of a light vector mediator $Z^{\prime}$ in the framework of the simplified universal model~\cite{Cerdeno_2016}, using the CE$\nu$NS detection channel.
Considering the lowest energy bin, 15--215\,eV, the expected event rate in CONNIE, $R_{SM+Z^{\prime}}$, is computed when accounting for the presence of the new mediator in addition to the SM interactions. 
The exclusion region in the phase space of the $Z^{\prime}$ mass ($M_{Z^{\prime}}$) and coupling ($g_{Z^{\prime}}$) is then calculated as the one in which this rate is greater than the observed upper limit on the event rate at 95\%~CL for the same energy bin.  

Figure~\ref{fig:LVlim} (left) shows the resulting CONNIE limits from the Skipper-CCD run with the updated Sarkis {\it et al.} quenching factor~\cite{Sarkis2023}. 
For illustration purposes, 
the limit corresponding to the Chavarria quenching factor is also given~\cite{Chavarria:2016xsi}, although recent measurements and calculations suggest it may be less valid at such low energies. 
The exclusion regions derived from the CONNIE Skipper-CCDs improve slightly the bounds imposed with the CONNIE standard CCDs using the Chavarria quenching factor~\cite{CONNIE2020}, mostly as an effect of the updated quenching factor at low energies, and to a lesser extent because the significantly higher Skipper-CCD rate uncertainty is compensated by the lower threshold, which enhances  sensitivity. 
A comparison with other experimental results in Fig.~\ref{fig:LVlim} (right) shows that the new CONNIE limit is less stringent but approaches that based on the COHERENT experiment CE$\nu$NS detection with CsI+Ar and its own quenching factor measurement~\cite{Corona:2022}. The limit is more stringent at low mediator masses compared to the CONUS experiment, which uses germanium detectors and the  Lindhard model with a quenching parameter of $k=0.16$~\cite{CONUS:2022}, thanks to the Skipper-CCD reach to lower energies.

Figure~\ref{fig:LVlim} (left) also shows the 95\%~CL upper limit projections assuming a zero difference between the reactor-on and off rates and a statistical error five times smaller in the lowest-energy bin.   
These limits may be achieved if CONNIE can collect 25 times more data by, for example, increasing its mass.

\subsection{Dark matter search by diurnal modulation}

\begin{figure*}[tb]
\centering
\includegraphics[width=0.49\textwidth]{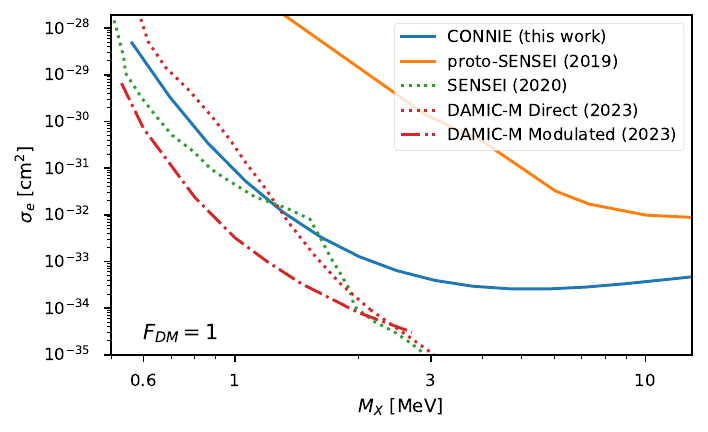}
\includegraphics[width=0.49\textwidth]{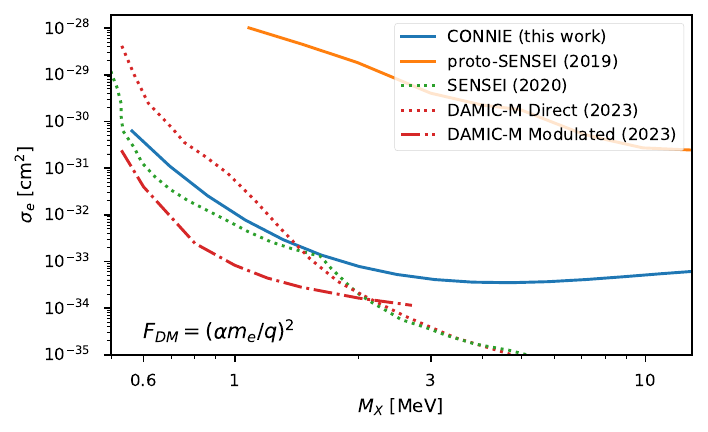}
\caption{90\% CL upper bounds on dark matter-electron interactions mediated by a heavy dark photon (left) and an ultralight dark photon (right) established by CONNIE (blue, solid), protoSENSEI \cite{SENSEI2018} (orange, solid), SENSEI \cite{sensei2020}(green, dotted), and DAMIC-M (red, dashed for modulation searches \cite{damicm2023mod} and red, dotted for direct searches~\cite{damicm2023}). Solid lines correspond to experiments running on the surface, dotted lines to direct limits obtained underground, and dashed to modulated limit obtained underground.} 
    \label{fig:DMlim}
\end{figure*}

Dark matter (DM) direct detection experiments using Skipper-CCDs have the best current constraints for light dark matter candidates interacting with electrons in most of the energy range of interest (0.5--1000\,MeV)~\cite{sensei2020, damicm2023}. 
These constraints are limited in the lowest-energy bin by the single-electron background rates. However, they can be improved by searching for a diurnal modulation, as explained in Ref.~\cite{bertoumod} and summarised below. 

Due to the Solar System movement around the galaxy, there is a preferred direction of arrival of dark matter particles, which is known as the DM wind. This wind comes, on average, from the latitude 40$^\circ$\,N\@. 
In the region of parameter space to which Skipper-CCDs are sensitive, DM particles interact with nuclei and electrons inside the Earth, and therefore can suffer elastic or inelastic collisions, producing a sidereal day modulation of their flux at the location of the detector. The ``isodetection angle'' is defined as the angle formed between the direction of the DM wind and the normal to the surface of the Earth at the detector location. Experiments located at the northern hemisphere scan low isodetection angles, meaning that the expected modulation is smaller (the DM particles only traverse the atmosphere and a small portion of the Earth crust in their trajectory). However, in the southern hemisphere, the isodetection angle spans larger angles and the expected modulation becomes bigger. CONNIE is located at 23$^\circ$\,S, allowing to scan a range of isodetection angles of roughly  [65--161]$^\circ$. 

Stable single-electron-event rates are needed in order to look for the modulation. To that end, we limited this analysis to the most stable data-taking run. Each image has an exposure of around one hour, and an average isodetection angle is assigned to it using the date and time information. The rates computed for each image are then binned in the isodetection angle. Simulations are carried out using DaMaSCUS~\cite{damascus1, damascus2} to calculate the expected rate for each isodetection angle bin. The simulations are then compared to the measured rates and a 90\% CL constraint for the DM parameters is calculated using a binned likelihood method.

The results are given in Fig.~\ref{fig:DMlim}, 
considering models with MeV-scale DM, which couple to SM particles via a kinetically mixed dark photon ($A^{\prime}$)~\cite{Galison:1983pa, Holdom:1985ag}.
The plots show the DM phase space limits in terms of the DM particle mass ($M_X$) and its interaction cross-section with electrons ($\sigma_e$). 
The DM-electron scattering cross-section depends on the DM form-factor, $F_{\rm DM} = (\alpha m_e/q)^n$, with $\alpha$ the fine structure constant, $m_e$ the electron mass and $q$ the momentum transfer, which varies with the mediator mass, $M_{A^{\prime}}$~\cite{Emken:2019tni}. 
For heavy dark photon mediators ($M_{A^{\prime}} \gg \alpha m_e$) $n=0$ and thus $F_{\rm DM} = 1$, while for ultralight mediators ($M_{A^{\prime}} \ll \alpha m_e$), $n=2$. 
The limits for the heavy and ultralight dark photon mediators are given in the left and right plots, respectively.

The figure also presents a comparison with other experiments. 
The plots in solid lines correspond to surface-level experiments, while dashed lines indicate underground experiments. 
In the [0.6--10]\,MeV region, this analysis establishes the best limits for a surface-level experiment by 1--3 orders of magnitude in the DM-electron cross-section. 
The CONNIE limits are not competitive with underground experiments due to the cosmogenic background, which introduces extra single-electron events produced by effects such as Cherenkov radiation in silicon~\cite{Du:2020ldo}. 
However, these are promising results for future searches in which a better stability of the detector is accomplished and the backgrounds are better understood.

\section{Summary and prospects}
\label{Sec:conclusion}

CONNIE has successfully installed and commissioned two Skipper-CCDs in its detector, becoming the first experiment to employ these novel sensors with the aim to detect reactor neutrinos. 
The experiment took data between November 2021 and December 2022, demonstrating stable operation and low noise, and collecting a total data exposure of 18.4\,g-days, after data quality selection. 
Despite the low exposure, resulting from the very low sensor total mass (0.25\,g), the experiment achieved excellent sensor performance, with an ultra-low readout noise of 0.150\,e$^-$, and a comparatively low single-electron rate, considering the surface location, of 0.045\,e$^-$/pix/day. 
This allowed us to greatly improve the detection efficiency reach at low energies, achieving a record low threshold among all CE$\nu$NS experiments of 15\,eV. 

The collected dataset was employed to look for rare processes, as a proof of principle for the possibilities of using Skipper-CCDs for searches within the SM and beyond. 
The sensitivity to CE$\nu$NS was extended to lower energies, where higher rates are expected, and the data quality selection was improved.
The resulting CE$\nu$NS limit, which is  dominated by the statistical uncertainties of the small dataset, it is comparable to our previous result,  obtained using an exposure greater by two orders of magnitude. 
The results demonstrate an increased sensitivity to probing CE$\nu$NS and new neutrino interactions at very low energies. 
For simplified models with new light vector mediators, the limit in the phase space of the model represents an improvement to our previous result.
A study of diurnal DM modulation was performed for the first time by CONNIE and the results represent the best limits on
the DM-electron scattering cross-section, obtained by a surface-level experiment, representing an improvement by 1–3 orders of magnitude. 
These first results using Skipper-CCD data  demonstrate the robustness and versatility of the sensors, as well as their promising potential to detect CE$\nu$NS and probe new physics models with high sensitivity.

In order to succeed in measuring these rare processes, CONNIE must increase its active sensor mass, while maintaining the quality and stability of the performance. 
Considering as an example a 1-kg detector with the current levels of readout noise, background rate and detection threshold, it would take only around a month to detect CE$\nu$NS at 90\% CL.  
The perspective for the immediate future is to increase the mass by installing 16 new sensors, mounted in a compact arrangement on a Multi-Chip-Module (MCM),  designed by Oscura~\cite{Oscura:2023qik}.
The plan is to assemble the MCM inside the current CONNIE detector setup by mounting it in place of the copper box of the standard and Skipper sensors. The rest of the detector design and shielding will be kept unchanged,
to ensure minimal disruption to the overall setup and a speedy commissioning. 
This upgrade is planned for the first half of 2024 and will bring a 32-fold increase in sensor mass to a total of 8\,g, thus greatly enhancing the physics potential of the CONNIE experiment.

\section*{Acknowledgements}

We thank the Silicon Detector Facility staff at the Fermi National Accelerator Laboratory for hosting the assembly and test of the detector components used in the CONNIE experiment. The CCD development was partly supported by the Office of Science, of the U.S. Department of Energy under Contract No. DE-AC02-05CH11231. 
We are grateful to Eletrobras Eletronuclear, and especially to Ilson Soares, Israel Ottoni Simas and Livia Werneck Oliveira, for access to the Angra 2 reactor site, infrastructure and the support of their personnel to the CONNIE activities. We express our gratitude to Ronald Shellard (in memoriam) for supporting the experiment. We thank Marcelo Giovani for his IT support. We are grateful to Sandra Amato for the useful discussions. 
We acknowledge the support from the Brazilian Ministry for Science, Technology, and Innovation and the Brazilian funding agencies FAPERJ (grants E-26/110.145/2013, E-26/210.151/2016, E-26/010.002216/2019, E-26/202.687/2019, E-26/210.079/2020), CNPq (grants 437353/2018-4, 407707/2021-2), and FINEP (RENAFAE grant 01.10.0462.00); Argentina's CONICET and AGENCIA I+D+i (grants PICT-2019-2019-04173; PICT-2021-GRF-TII-00458; PICT-2021-GRF-TI-00816) and Mexico’s CONAHCYT (grant No. CF-2023-I-1169) and DGAPA-UNAM (PAPIIT grant IN104723). The UNAM group thanks Ing.~Mauricio Martínez Montero for his technical assistance in skipper configuration studies.
IN is also supported by a Latin American Association for High Energy, Cosmology and Astroparticle Physics Network (LAA-HECAP Network) grant through the ICTP Network Program.
This work made use of the CHE cluster, managed and funded by COSMO/CBPF/MCTI, with financial support from FINEP and FAPERJ, and operating at the Javier Magnin Computing Center/CBPF.

\appendix

\section{Observed rates and expected neutrino flux}
\label{AppendixRates}

\begin{table}[tb]
\centering
\renewcommand{\arraystretch}{1.5}
\begin{tabular}{c|c|c|c}
Energy range & Reactor-off   &  Reactor-on  & Difference \\ 
         
[keV] & rate [kdru] & rate [kdru] & [kdru] \\ 
         \hline
$0.015 - 0.215$ & $2.9_{-2.9}^{+2.2}$ & $2.0_{-0.9}^{+0.8}$ & $-0.9_{-3.0}^{+2.3}$ \\ 
$0.215 - 0.415$ & $1.4_{-1.4}^{+2.0}$ & $4.0_{-1.2}^{+1.2}$ & $+2.6_{-1.8}^{+2.3}$ \\ 
$0.415 - 0.615$ & $2.9_{-2.9}^{+2.2}$ & $3.0_{-1.1}^{+1.0}$ & $+0.1_{-3.1}^{+2.4}$ \\ 
$0.615 - 0.815$ & $5.7_{-3.3}^{+2.9}$ & $2.7_{-1.0}^{+0.9}$ & $-3.0_{-3.5}^{+3.0}$ \\ 
$0.815 - 1.015$ & $0.0_{-0.0}^{+1.7}$ & $4.0_{-1.2}^{+1.2}$ & $+4.0_{-1.2}^{+2.0}$ \\ 
$1.015 - 1.215$ & $4.3_{-3.1}^{+2.5}$ & $4.4_{-1.2}^{+1.2}$ & $+0.1_{-3.3}^{+2.8}$ \\ 
$1.215 - 1.415$ & $2.9_{-2.9}^{+2.2}$ & $2.7_{-1.0}^{+0.9}$ & $-0.2_{-3.1}^{+2.4}$ \\ 
$1.415 - 1.615$ & $1.4_{-1.4}^{+2.0}$ & $4.4_{-1.2}^{+1.2}$ & $+3.0_{-1.9}^{+2.3}$ \\ 
$1.615 - 1.815$ & $13.0_{-4.6}^{+4.3}$ & $7.7_{-1.6}^{+1.6}$ & $-5.3_{-4.9}^{+4.6}$ \\ 
$1.815 - 2.015$ & $1.4_{-1.4}^{+2.0}$ & $2.0_{-0.9}^{+0.8}$ & $+0.6_{-1.7}^{+2.1}$ \\ 
\end{tabular}
\caption{Observed rates with CONNIE Skipper-CCD reactor-on and reactor-off data, and the difference between the two.}
\label{tab:observedlimit}
\end{table}

\begin{figure}[tb]
\centering
\includegraphics[width=0.9\linewidth]{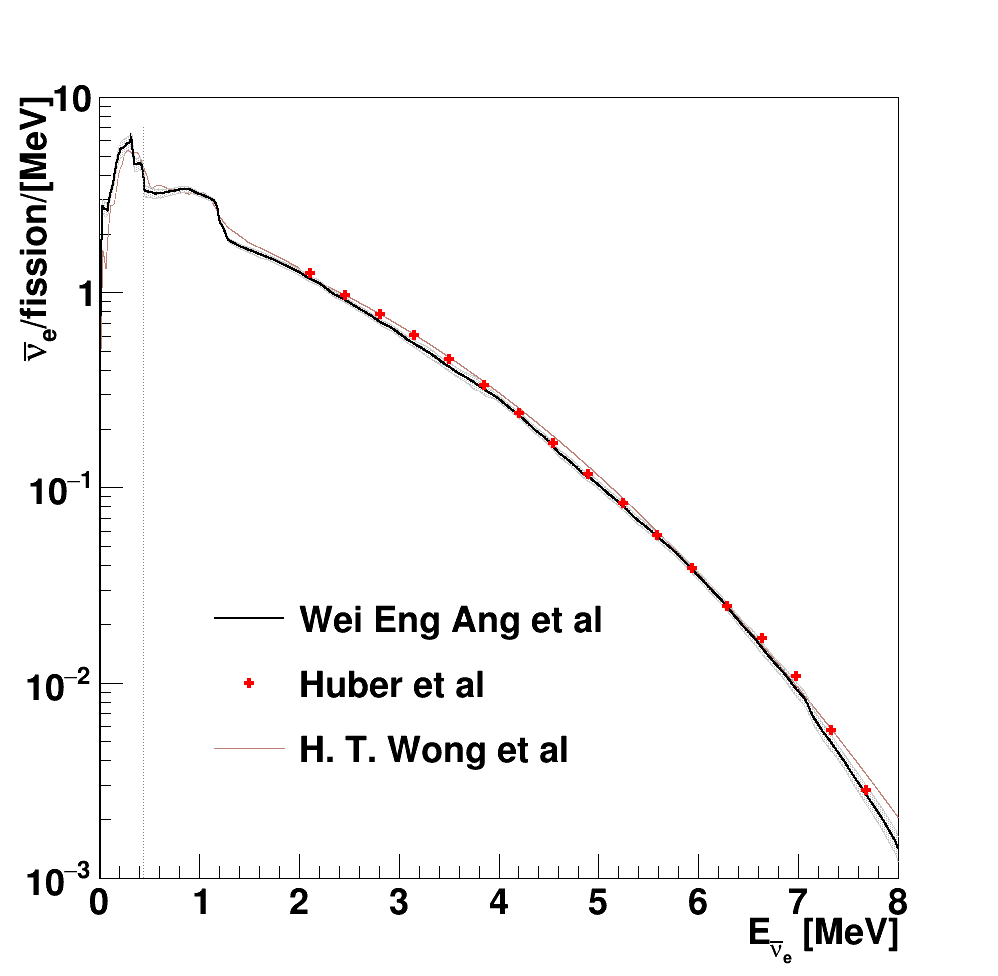}
\caption{Total reactor antineutrino spectrum per
fission in the reactor per MeV for three models: (black) Wei Eng Ang {\it et al.}~\cite{NufluxSumationMeth}, (red) Huber {\it et al.}~\cite{HuberSpectrum} and (brown) Wong~\cite{Oldflux}. The error band for Wei model is computed according to~\cite{PerisePWRErr,ReactorNuSumationPerise2}. The dotted vertical line denotes the neutrino energy threshold for Skipper-CCDs at CONNIE.}
    \label{fig:Nuspec}
\end{figure}

Table~\ref{tab:observedlimit} shows the observed rates in bins of energy for data with the reactor on and off, and the difference between the two. The numbers correspond to the plot of Figure~\ref{fig:SpectrumCONNIEskp}.

Figure~\ref{fig:Nuspec} shows the updated antineutrino flux at the detector. 
It was calculated following the summation method described in Refs.~\cite{NufluxSumationMeth,ReactorNuSumationPerise,ReactorNuSumationPerise2}, updated with improved antineutrino spectra for the fissile isotopes $^{235}\rm{U}$, $^{238}\rm{U}$, $^{239}\rm{Pu}$ and $^{241}\rm{Pu}$, as well as with the contribution from activation (neutron capture) of structural and fuel elements relevant for energies below 1.27 MeV following Ref.~\cite{FluxNuT}. 
The threshold of 15\,$\rm{eV_{ee}}$ corresponds to a minimum observable neutrino energy of $\sim$0.44~MeV, shown in the figure, according to the Sarkis quenching factor model~\cite{Sarkis2023}. Above this value, the updated flux model and the one used previously \cite{ChavarriaQF} agree to within 3\% and are thus considered equivalent. 
The main difference between the new and old flux models appears for $E_{\nu}\lesssim$ 0.35~MeV, which is below the minimum observable neutrino energy.

\begin{figure*}[tb]
\centering
\includegraphics[scale=0.22]{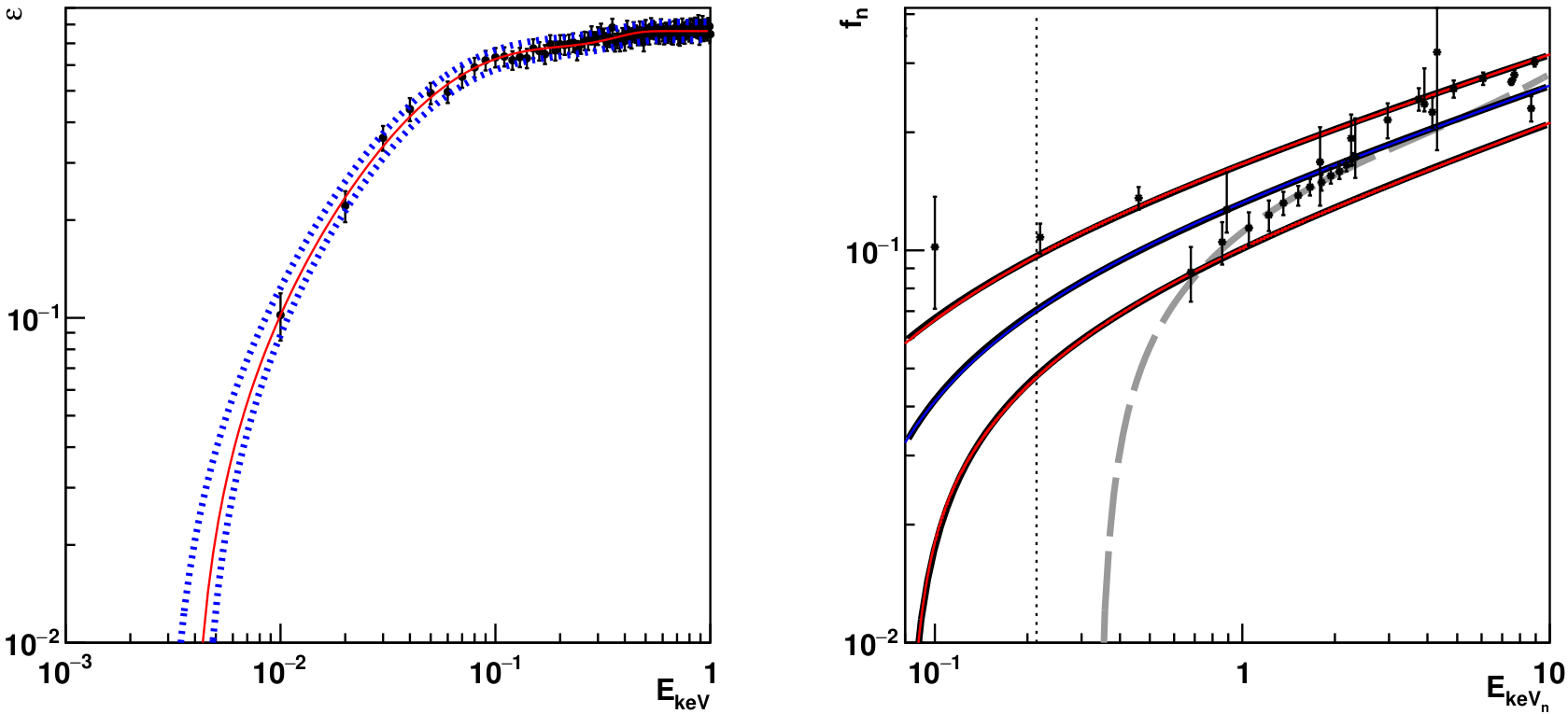}
\caption{(Left) Efficiency points with error bars compared with the fit function from Eq.~\ref{EffFit} (red). (Right) Ionisation efficiency from \cite{Sarkis2023} (blue),  for the upper and lower curves (red) compared with the fit function from Eq.~\ref{QFFit} (black) and from Chavarria \cite{ChavarriaQF} (long dash). The vertical dotted line shows the threshold recoil energy ($0.24\,\rm{keV_n}$) for Skipper-CCDs using Eq.~\ref{QFFit}.  Also shown are the measurements extracted from Refs.~\cite{Sarkis2023,CDMSQF}.}
\label{fig:QFEffCompare}
\end{figure*}

\section{Fitting functions for detection and ionisation efficiency}
\label{Appendix1}

To obtain and reconstruct the events registered during a Skipper-CCD exposure, a specific set
of processing tools are used. The reconstruction efficiency for these tools has been evaluated using simulated events as discussed in Section \ref{sec:Efficiency}. This efficiency can be fitted by the continuous function,
\begin{equation}\label{EffFit}
     \varepsilon(E_I)=b_0(1-b_1\exp(-b_2(E_I)^4) -b_3\exp(-b_4E_I)),
 \end{equation}
where $b_0=0.763599$, $b_1=0.108047$, $b2=51.1075$, $b_3=0.984304$ and $b_4=26.0671$ for the central curve only, with a $\chi^2=20/95$, see fig. \ref{fig:QFEffCompare}. For the upper efficiency curve the parameters are, $b_0=0.813653$, $b_1=0.115169$, $b_2=19.5823$, $b_3=0.955748$ and $b_4=26.5726$. And for the lower efficiency curve the parameters are, $b_0=0.717902$, $b_1=0.122821$, $b_2=19.9976$, $b_3=0.980065$ and $b_4=26.2413$.\\

For the ionisation efficiency we used the first principles model from Ref.~\cite{Sarkis2023}, where we fit the continuous function,
\begin{equation}\label{QFFit}
     f_n(E_R)=\frac{( a_1(E_R-E_0) +a_2(E_R-E_0)^{a_3} +a_4(E_R-E_0)^{a_5} )}{(a_6(E_R-E_0) +a_7(E_R-E_0)^{a_8})},
\end{equation}
from $E_R=0.08$ keV to $E_R=10.00$ keV such that the relative error from the exact numeric tabulated value is less than 1\%, see Fig.~\ref{fig:QFEffCompare}.
The fit parameters for the central curve are, $E_0=0.0386870 \; (\rm{keV}) ,a_1=1.35347 \;(1/\rm{keV}), a_2=-0.794527 \; (1/\rm{keV^{a_3}}), a_3=0.982405, a_4=-0.461248\; (1/\rm{keV^{a_5}}), a_5=0.987530, a_6=-0.911767 \; (1/\rm{keV}), a_7=1.64321 \; (1/\rm{keV^{a_8}})$, and $a_8=0.946677$.\\

For the upper curve the fit parameters are, $E_0=0.0152147 \; (\rm{keV})$, $a_1=0.205447 \;(1/\rm{keV})$, $a_2=-0.0122554 \; (1/\rm{keV^{a_3}})$, $a_3=0.709046$ , $a_4=-0.0274662\; (1/\rm{keV^{a_5}})$, $a_5=0.649680$, $a_6=-0.0331458 \; (1/\rm{keV})$, $a_7=1.02803 \; (1/\rm{keV^{a_8}})$, and $a_8=0.784991$. 
For the lower curve the fit parameters are, $E_0=0.0724154 \; (\rm{keV})$, $a_1=0.526887 \;(1/\rm{keV})$, $a_2=0.00140378 \; (1/\rm{keV^{a_3}})$, $a_3=0.283309$, $a_4=-0.241916\; (1/\rm{keV^{a_5}})$, $a_5=0.828205$, $a_6=-0.289727 \; (1/\rm{keV})$, $a_7=3.05516 \; (1/\rm{keV^{a_8}})$, and $a_8=0.819846$.


\bibliography{main}

\end{document}